\documentclass[conference]{IEEEtran}
\pdfoutput=1
\usepackage[pdftex]{graphicx}
\DeclareGraphicsExtensions{.pdf,.jpeg,.png}
\usepackage[cmex10]{amsmath}
\usepackage{commath}
\usepackage{amsfonts, amssymb}
\usepackage{esint}
\usepackage{hyperref}
\usepackage{multirow}
\newcommand{\kf}{k_f} 
\newcommand{\kb}{k_b}
\newcommand{\kd}{k_d} 
\newcommand{\pr}{\text{Pr}}

\newcommand{\w}{\text{W}}
\newcommand{\erf}{\mathrm{erf}\,}
\newcommand{\erfc}{\mathrm{erfc}\,} 
\newcommand{\pacv}[1]{P_{AC}({#1}|\vec{r}_0)}
\newcommand{\pac}[1]{P_{AC}({#1}|r_0)}
\newcommand{\pav}[2]{P_{A}({#1},{#2}|\vec{r}_0)}
\newcommand{\pa}[2]{P_{A}({#1},{#2}|r_0)}
\newcommand{\hs}{\hspace*{-0.75 mm}}
\allowdisplaybreaks
\begin{document}

\title{Reactive Receiver Modeling for Diffusive Molecular Communication Systems with Molecule Degradation}
\author{ 
\IEEEauthorblockN{
Arman Ahmadzadeh\IEEEauthorrefmark{1},
Hamidreza Arjmanidi\IEEEauthorrefmark{2}, 
Andreas Burkovski\IEEEauthorrefmark{1}, 
and Robert Schober\IEEEauthorrefmark{1}
} 
\IEEEauthorblockA{\IEEEauthorrefmark{1}Friedrich-Alexander-Universit{\"a}t Erlangen-N{\"u}rnberg, Germany \, \IEEEauthorrefmark{2}Sharif University of Thechnology, Iran}
}
\maketitle 
\begin{abstract} 
In this paper, we consider the diffusive molecular communication channel between a transmitter nano-machine and a receiver nano-machine in a fluid environment. The information molecules released by the transmitter nano-machine into the environment can degrade in the channel via a first-order degradation reaction and those that reach the receiver nano-machine can participate in a reversible bimolecular-reaction with receiver receptor proteins. We derive a closed-form analytical expression for the expected received signal at the receiver, i.e., the expected number of activated receptors on the surface of the receiver. The accuracy of the derived analytical result is verified with a Brownian motion particle-based simulation of the environment.          
\end{abstract}
\section{Introduction} 
In nature, one of the primary means of communication among biological entities, ranging from organelles to organisms, is molecular communication (MC), where molecules are the carriers of information. MC is also an attractive option for the design of intelligent man-made communication systems at nano and micro scale. Sophisticated man-made MC systems are expected to have various biomedical, environmental, and industrial applications \cite{NakanoB}.    

Similar to any other communication system, the design of basic functionalities such as modulation, detection, and estimation requires accurate models for the transmitter, channel, and receiver. However, the modeling of these components in MC systems is vastly different from the modeling of traditional communication systems as the size of the nodes of MC systems is on the order of tens of nanometer to tens of micrometer \cite{NakanoB}. 
Furthermore, in MC systems, the communication environment (or channel) and the impairments occurring in the channel are completely different from those in traditional communication systems. For example, in a typical MC setup, the transmitter and the receiver are embedded in a \emph{fluid} environment, where the transmitter employs a certain type of molecule to communicate with the receiver. 
Additionally, the information-carrying molecules may be affected by different environmental effects such as \emph{flow} and/or \emph{chemical reactions}, which may prevent them from reaching the receiver \cite{NakanoB}.

The molecules that succeed in reaching the receiver may react with the receptors on the surface of the receiver and activate them. These activated receptors can be interpreted as the received signal. This is a common reception mechanism in natural biological cells. In particular, cells measure the presence of information-carrying molecules via receptors on their surface \cite{AlbertsBook}. These measurements are inevitably corrupted by noise that arises from the stochastic arrival of the molecules by diffusion and from the stochastic binding of the molecules to the receptors. Once an information-carrying molecule binds to a receptor, i.e., \emph{activates the receptor}, this receptor transduces the received noisy message into a cell response via a set of signaling pathways, see \cite[Chapter 16]{AlbertsBook}. Thus, having a meaningful model for the reception process that captures the effects of the main phenomena occurring in the channel and at the receiver is of particular importance and is the focus of this paper.

The modeling of the \emph{reception mechanism} at the receiver has been studied in the existing MC literature; see \cite{PierobonJ1, PierobonJ2, MahfuzJ1, PierobonJ3, ShahMohammadianProc1, YilmazL1, ChunJ1}. Most of these works assume that the receiver is transparent and the received signal is approximated by the local concentration of the information-carrying molecules inside the receiver, see \cite{PierobonJ1, PierobonJ2, MahfuzJ1}. However, this approach neglects the impact of the chemical reactions at the receiver surface required for sensing the concentration of molecules in its vicinity. In a first attempt to consider the effect of chemical reactions at the receiver, the authors in \cite{PierobonJ3} and \cite{ShahMohammadianProc1} modelled the reception mechanism at the receiver using the theory of ligand-receptor binding kinetics, where molecules (ligands) released by a transmitter can reversibly react with receptors at the receiver surface and produce output molecules whose concentration is then referred to as the output signal in \cite{PierobonJ3}, \cite{ShahMohammadianProc1}. Analytical expressions that relate the output signal to the concentration of molecules at the receiver were derived. However, in \cite{PierobonJ3} and \cite{ShahMohammadianProc1}, the authors assumed the diffusion of molecules to be independent from the reaction mechanisms at the receiver, i.e., the equations describing the output signal were derived for a given concentration at the receiver. This assumption may not be justified as the reaction-diffusion equation is highly \emph{coupled} and cannot be separated. The authors in \cite{YilmazL1} approximated the reception mechanism by modeling the receiver as a fully-absorbing sphere, where molecules released by the transmitter are absorbed by the receiver as soon as they hit its surface, and derived an analytical time domain expression for the number of absorbed molecules. However, it may not be realistic to assume that every collision of a molecule with the receiver surface triggers a reaction. In the recent work \cite{ChunJ1}, the reception mechanism was approximated by a first-order reversible reaction inside the receiver, and a reaction-diffusion master equation (RDME) model was used to solve the corresponding coupled reaction-diffusion equation. An analytical expression for the expected number of output molecules was derived in the Laplace domain. However, no time-domain solution was provided.

In this paper, we model the reception mechanism at the receiver surface as a second-order reversible reaction, where an information molecule released by the transmitter can \emph{reversibly} react with receptor protein molecules covering the surface of the receiver and activate a receptor protein molecule, which can later be used for detection. Furthermore, unlike \cite{PierobonJ1, PierobonJ2, MahfuzJ1, PierobonJ3, ShahMohammadianProc1, YilmazL1, ChunJ1}, we also take into account that in a realistic environment, molecules released by the transmitter may degrade in the channel via a first-order degradation reaction. We derive a novel closed-form time domain expression for the expected number of activated receptor protein molecules in response to the impulsive release of molecules at the transmitter. We verify the accuracy of the derived analytical expression via stochastic simulation of the environment using a Brownian motion particle-based approach.     

The rest of this paper is organized as follows. In Section \ref{Sec.SysMod}, we introduce the system model. In Section \ref{Sec.CIR}, we derive a closed-form analytical expression for the expected number of activated receptor protein molecules which we refer to as the received signal. In Section \ref{Sec. Simulations}, we briefly describe the framework used for the adopted particle-based simulator and present simulation and analytical results. Finally, some conclusions are drawn in Section \ref{Sec.Con}.   
\section{system model} 
\label{Sec.SysMod} 
In this section, we introduce the system model considered in this paper. 
\begin{figure}[!t] 
	\centering
	\includegraphics[scale = 1.7]{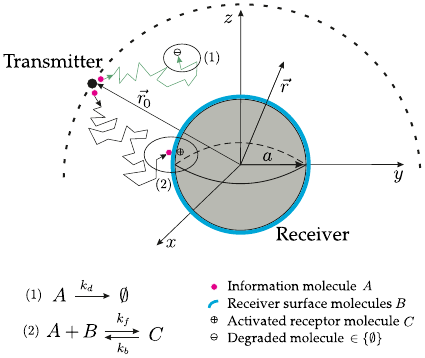}\vspace*{-2 mm}
	\caption{Schematic diagram of the considered system model, where receiver and transmitter are shown as a gray sphere in the center of the coordinate system and a black dot, respectively. The surface of the receiver is uniformly covered with type $B$ molecules, shown in blue color. Two sample trajectories of an $A$ information molecule (denoted by red dots) that result in a degraded molecule (circle with a ``-'' inside) and an activated receptor molecule $C$ (circle with a ``+'' inside ) are shown as black and green arrows, respectively.}\vspace*{-5 mm} 
	\label{Fig.SystemModel}
\end{figure} 
We consider an unbounded three-dimensional fluid environment with constant temperature and viscosity. We assume that the transmitter is a point source placed at position $\vec{r}_0 = (x_0,\ y_0,\ z_0)$ and the receiver is spherical in shape with a fixed radius $a$ and placed at the center of the coordinate system, see Fig. \ref{Fig.SystemModel}. 

We assume that the transmitter uses one specific type of molecule, denoted by $A$, for sending information to the receiver. Hence, we refer to the information molecules as $A$ molecules. Furthermore, we assume that the transmitter instantaneously releases a fixed number of $A$ molecules, $N_A$, at time $t_0 = 0$ into the environment. 

After their release, the information molecules diffuse in the environment into all directions with constant diffusion coefficient $D_A$. Thereby, we assume that the diffusion processes of different information molecules are independent of each other. We furthermore assume that the $A$ molecules can be degraded throughout the environment via a reaction mechanism of the form 
\begin{equation}
	\label{Eq.DegReaction} 
	A \xrightarrow{\kd} \emptyset,  
\end{equation}
where $\kd$ is the degradation reaction constant in $\text{s}^{-1}$, and $\emptyset$ is a species of molecules which is not recognized by the receiver. Eq. (\ref{Eq.DegReaction}) models a first-order reaction but can also be used to approximate higher order reactions, see e.g. \cite{Steven_Andrews, NoelJ1, ChouJ1}. 

Some of the $A$ molecules released by the transmitter may reach the receiver surface, which we assume to be uniformly covered by receptor protein molecules of type $B$. An $A$ molecule can \emph{reversibly} react with a $B$ molecule to form an activated receptor (also referred to as ligand-receptor complex molecule), denoted by $C$, via a reaction mechanism of the form 
\begin{equation}
	\label{Eq.RevReaction} 
	A + B \mathrel{\mathop{\rightleftarrows}^{\kf}_{\kb}} C,  
\end{equation}
where $\kf$ and $\kb$ are the microscopic forward reaction constant in $\text{molecule}^{3}\cdot \text{s}^{-1}$ and the microscopic backward reaction constant in $\text{s}^{-1}$, respectively. In nature, different cell-surface receptors produce different responses inside the cell via different signaling pathways once they are activated \cite[Chapter 16]{AlbertsBook}. However, in this work, we do not focus on a specific signaling pathway. Instead, we consider the formation of the $C$ ligand-receptor complex molecules as the \emph{received signal} at the receiver, which could be used for detection of information sent by the transmitter. Furthermore, in order to make our analysis analytically tractable, we adopt the following assumptions in \eqref{Eq.RevReaction}:
\begin{itemize}
	\item We do not model an individual receptor but assume that the \emph{whole} surface of the receiver is covered with sufficiently many receptor protein molecules $B$.
	\item We neglect the effect of receptor saturation, which means that multiple $A$ molecules can react at the same position on the surface of the receiver to form multiple $C$ molecules.  
\end{itemize}
With these two assumptions the formations of different $C$ molecules on the surface of the receiver become independent of each other.  
\section{expected received signal}
\label{Sec.CIR} 
In this section, we first formulate the problem that has to be solved to find the expected received signal at the receiver for the considered MC system. Subsequently, we derive a closed-form expression for the probability of finding an $A$ molecule, which can undergo reactions (\ref{Eq.DegReaction}) and (\ref{Eq.RevReaction}), at the position defined by vector $\vec{r}$ at time $t$, given that it was released at position $\vec{r}_0$ at time $t_0$. Finally, using this result, we derive a closed-form expression for the channel impulse response.

\subsection{Problem Formulation}
We define the \emph{expected} received signal at the receiver at time $t$, $\overline{N}_C(t)$, as the time-varying number of $C$ molecules expected on the surface of the receiver at time $t$ when the transmitter released at time $t_0$ an impulse of $N_A$ $A$ molecules into the environment. This signal is a function of time, due to the random walks of the molecules. In this subsection, we formulate the problem that has to be solved to find $\overline{N}_C(t)$. Considering the assumption of independent movement of individual molecules, we have $\overline{N}_C(t) = N_A \pacv{t}$, where $\pacv{t}$ is the probability that a given $A$ molecule, released at $\vec{r}_0$ and time $t_0=0$, causes the formation of an activated receptor molecule $C$ on the surface the receiver at time $t$. We refer to this probability also as the \emph{channel impulse response} of the system. 

Furthermore, the probability that a given $A$ molecule released at $\vec{r}_0$ at time $t_0=0$ is at position $\vec{r}$ at time $t$, given that this molecule may undergo either of the two reaction introduced in (\ref{Eq.DegReaction}) and (\ref{Eq.RevReaction}) during time $t$, is denoted by $\pav{\vec{r}}{t}$. Assuming for the moment $\pav{\vec{r}}{t}$ is known, we can evaluate the incoming probability flux\footnote{We note that the probability flux refers to the flux of the position probability of a \emph{single} $A$ molecule, whereas the conventional diffusive molecular flux (used in Fick's first law of diffusion) refers to the flux of the average number of $A$ molecules. For further details, we refer the interested reader to \cite[Chapter 3]{GillespieBook}.}, $-J(\vec{r},t | \vec{r}_0)$, at the surface of the receiver by applying Einstein's theory of diffusion as \cite[Eq. (3.34)]{GillespieBook} 
\begin{equation}
	\label{Eq. ProbFlux} 
	-J(\vec{r},t | \vec{r}_0)\big|_{\vec{r} \in \Omega} = D_A \nabla \pav{\vec{r}}{t}\big|_{\vec{r} \in \Omega},
\end{equation}
where $\nabla$ is the gradient operator in spherical coordinates and $\Omega$ is the surface of the receiver. Now, given $-J(\vec{r},t | \vec{r}_0)$, $-J(\vec{r},t | \vec{r}_0) \dif  \Omega \dif t$ is the probability that a given $A$ molecule reacts with the infinitesimally small surface element $\dif \Omega$ of the receiver during infinitesimally small time $\dif t$. Integrating this function over time and the surface of the receiver yields a relationship between $\pacv{t}$ and $\pav{\vec{r}}{t}$, which can be written as \cite[Eq. (3.35)]{GillespieBook} 
\begin{equation}
	\label{Eq. Relationship} 
	\pacv{t} = - \int_{0}^{t} \oiint_{\Omega} J(\vec{r},t^\prime | \vec{r}_0)\cdot \dif \Omega \dif t^\prime ,   
\end{equation} 
Thus, for evaluation of $\pacv{t}$, we first need to find $\pav{\vec{r}}{t}$.  

Clearly, the evaluation of (\ref{Eq. Relationship}), and subsequently (\ref{Eq. ProbFlux}), requires knowledge of $J(\vec{r},t | \vec{r}_0)$ only on the surface of the receiver, i.e., on $\Omega$. However, since we assume that the surface of the receiver is uniformly covered by type $B$ molecules, $J(\vec{r},t | \vec{r}_0)$ and $\pav{\vec{r}}{t}$ are only functions of the magnitude of $\vec{r}$, denoted by $r$, and not of $\vec{r}$ itself. This can also be intuitively understood from Fig. \ref{Fig.SystemModel}. To this end, let us assume that the point source transmitter is mounted on top of a virtual sphere with radius $r_0$, where $r_0$ is the magnitude of $\vec{r}_0$. Because of symmetry, we expect that releasing a given molecule $A$ from any arbitrary point on the surface of the virtual sphere leads to the same probability of reaction on the surface of the receiver, i.e., the same $\pacv{t}$ results. However, this is only possible if $J(\vec{r},t | \vec{r}_0)$ in (\ref{Eq. Relationship}) is independent of azimuthal angle $\theta$ and polar angle $\phi$ and only depends on $r$. In the remainder of this paper, we substitute $\vec{r}$ with its magnitude $r$ in (\ref{Eq. ProbFlux}) and (\ref{Eq. Relationship}) without loss of generality. As a result, (\ref{Eq. ProbFlux}) and (\ref{Eq. Relationship}) can be combined as 
\begin{equation}
	\label{Eq. RelationshipReduced} 
	\pac{t} = \int_{0}^{t} 4\pi a^2 D_A \frac{\partial \pa{r}{t^\prime}}{\partial r} \bigg |_{r = a} \dif t^\prime.    
\end{equation}  

\newcounter{tempequationcounter} 
\begin{figure*}[!t]
\normalsize
\setcounter{tempequationcounter}{\value{equation}}
\begin{IEEEeqnarray}{rCl}
	\setcounter{equation}{16}
	\label{Eq. Green's function}
	\pa{r}{t} & = & \exp(-k_d t) \Biggl[ \frac{1}{8 \pi r r_0 \sqrt{\pi D_A t} } 
	 \left( \exp\left(\frac{-\left( r - r_0 \right)^2}{4 D_A t}\right)  
	 + \exp\left( \frac{-\left( r + r_0 - 2a \right)^2}{4D_A t} \right) \right) - \frac{1}{4 \pi r r_0 \sqrt{D_A}}  \nonumber \\
	&& \times \> \left( \eta_1 \w \left( \frac{r + r_0 - 2a}{\sqrt{4 D_A t}}, \alpha \sqrt{t} \right) 
	 + \eta_2 \w \left( \frac{r + r_0 - 2a}{\sqrt{4 D_A t}}, \beta \sqrt{t}  \right) 
	 + \eta_3 \w \left( \frac{r + r_0 - 2a}{\sqrt{4 D_A t}}, \gamma \sqrt{t}  \right) \right)  \Biggr].
\end{IEEEeqnarray}
\setcounter{equation}{\value{tempequationcounter}}
\hrulefill
\vspace*{4pt} 
\vspace*{-6 mm}
\end{figure*}

In the following, we are interested in formulating the problem of finding $\pa{r}{t}$ for the system model specified in Section \ref{Sec.SysMod}. In order to do so, we start with the general form of the reaction-diffusion equation for the degradation reaction in (\ref{Eq.DegReaction}), which can be written as \cite{GrinrodB1} 
\begin{equation}
	\label{Eq. Reaction-Diffusion} 
	\frac{\partial \pa{r}{t}}{\partial t} = D_A \nabla^{2} \pa{r}{t} - k_d \pa{r}{t}, 
\end{equation}
where $\nabla^{2}$ is the Laplace operator in spherical coordinates. As discussed above, releasing a given $A$ molecule from any point on the surface of a sphere with radius $r_0$ including the point defined by $\vec{r}_0$, i.e., the actual position of the transmitter, results in the same channel impulse response, $\pac{t}$. Thus, a point source defined by $\vec{r}_0$ and a source uniformly distributed on the sphere with radius $r_0$ are equivalent in this context. As a result, releasing a given $A$ molecule at $\vec{r}_0$ at time $t_0 = 0$ can be modelled with the following initial condition 
\begin{equation}
	\label{Eq. InitialCondition} 
	\pa{r}{t \to 0} = \frac{1}{4\pi r_{0}^{2}} \delta(r - r_0),
\end{equation}
where constant $1/(4\pi r_{0}^{2})$ is a normalization factor and $\delta(\cdot)$ is the Dirac delta function. The boundary conditions of the system model for the assumed unbounded environment and the reaction mechanism in (\ref{Eq.RevReaction}) on the surface of the receiver can be written as \cite[Eqs. (3), (4)]{H.KimJ1} 
\begin{equation}
	\label{Eq. BounCon1}
	\lim_{r \to \infty} \pa{r}{t} = 0,
\end{equation} 
and 
\begin{IEEEeqnarray}{rCl}
	\label{Eq. BounCon2} 
	4\pi a^2 D_A \frac{\partial \pa{r}{t}}{\partial r} \bigg |_{r = a} & \hspace*{-1 mm} = & k_f \pa{a}{t} 	 
	 - k_b  [1 - S(t|r_0)], \nonumber \\*    
\end{IEEEeqnarray}
respectively, where $S(t|r_0)$ is the probability that a given $A$ molecule, released at distance $r_0$ at time $t_0$, has neither reacted on the boundary of the receiver nor degraded in the channel by time $t$. $S(t|r_0)$ can be obtained as follows: 
\begin{IEEEeqnarray}{c}
	\label{Eq. SurvivingProb} 
	S(t|r_0) = 1 - \int_{0}^{t} 4\pi a^2 D_A \frac{\partial \pa{r}{t^\prime}}{\partial r} \bigg |_{r = a} \dif t^\prime .  
\end{IEEEeqnarray}

The solution of reaction-diffusion equation (\ref{Eq. Reaction-Diffusion}) with initial and boundary conditions (\ref{Eq. InitialCondition})-(\ref{Eq. BounCon2}) is the Green's function\footnote{The solution of an inhomogeneous partial differential equation with initial condition in a form of Dirac delta function is referred to as the Green's function \cite{Ivar}.} of our system model. 

To summarize, obtaining the expected received signal $\overline{N}_C(t)$ requires knowledge of the impulse response $\pac{t}$ which, in turn, can be found via the Green's function based on (\ref{Eq. RelationshipReduced}).
\subsection{Green's Function} 
In this subsection, we derive a closed-form analytical expression for the Green's function of the system. To this end, we adopt the methodology introduced in \cite{SchultenL1}. In particular, we decompose $\pa{r}{t}$ as 
\begin{IEEEeqnarray}{c}
	\label{Eq. Partition} 
	\pa{r}{t} = U(r,t|r_0) + V(r,t | r_0),
\end{IEEEeqnarray}
where function $U(r,t|r_0)$ is chosen such that it satisfies both the reaction-diffusion equation (\ref{Eq. Reaction-Diffusion}) and initial condition (\ref{Eq. InitialCondition}). On the other hand, function $V(r,t | r_0)$ is chosen such that it satisfies (\ref{Eq. Reaction-Diffusion}), but at the same time satisfies jointly with function $U(r,t | r_0)$ the boundary conditions (\ref{Eq. BounCon1}) and (\ref{Eq. BounCon2}). With this approach, we can decompose the original problem into two sub-problems as follows. In the first sub-problem, we solve the reaction-diffusion equation 
\begin{IEEEeqnarray}{c}
	\label{Eq. Reaction-DiffusionU} 
	\frac{\partial U(r,t | r_0)}{\partial t} = D_A \nabla^{2} U(r,t | r_0) - k_d U(r,t | r_0), 
\end{IEEEeqnarray} 
with initial condition 
\begin{IEEEeqnarray}{c}
	\label{Eq. InitialConditionU} 
	U(r,t \to 0 | r_0) = \frac{1}{4\pi r_{0}^{2}} \delta(r - r_0).
\end{IEEEeqnarray}
In the second sub-problem, we solve the reaction-diffusion equation
\begin{IEEEeqnarray}{c}
	\label{Eq. Reaction-DiffusionV} 
	\frac{\partial V(r,t | r_0)}{\partial t} = D_A \nabla^{2} V(r,t | r_0) - k_d V(r,t | r_0), 
\end{IEEEeqnarray}
with the initial condition 
\begin{IEEEeqnarray}{c}
	\label{Eq. ConditionV} 
	V(r,t \to 0 | r_0) = 0.
\end{IEEEeqnarray} 
Finally, the solutions of both sub-problems are combined, cf. (\ref{Eq. Partition}), such that they jointly satisfy boundary conditions (\ref{Eq. BounCon1}) and (\ref{Eq. BounCon2}). The final solution for the Green's function is given in the following theorem. 

\newtheorem{theorem}{Theorem}
\begin{theorem}[Green's function]
The probability of finding a given $A$ molecule at distance $r \geq r_0$ from the center of the receiver at time $t$, given that it was released at distance $r_0$ at time $t_0 = 0$ and may be degraded via first-order degradation reaction  (\ref{Eq.DegReaction}) and/or react with the receptor molecules at the receiver surface via the second-order reversible reaction (\ref{Eq.RevReaction}) during time $t$ is given by \eqref{Eq. Green's function} at the top of this page,\addtocounter{equation}{1}
where function $\w(n,m)$ is defined as 
\begin{IEEEeqnarray}{c} 
	\label{Eq. FunctionW} 
	\w(n,m) = \exp\left( 2nm + m^2 \right) \erfc\left( n + m \right),  
\end{IEEEeqnarray} 
$\erfc(\cdot)$ is the complementary error function, and constants $\eta_1$, $\eta_2$, and $\eta_3$ are given by 
\begin{IEEEeqnarray}{rCl} 
	\label{Eq. eta1} 
	\eta_1 & = & \frac{\alpha(\gamma + \alpha)(\alpha + \beta)}{(\gamma - \alpha)(\alpha - \beta)}, \\
	\label{Eq. eta2} 
	\eta_2 & = & \frac{\beta(\gamma + \beta)(\alpha + \beta)}{(\beta - \gamma)(\alpha - \beta)}, \\
	\label{Eq. eta3} 
	\eta_3 & = & \frac{\gamma(\gamma + \beta)(\alpha + \gamma)}{(\beta - \gamma)(\gamma - \alpha)},
\end{IEEEeqnarray}
respectively. Here, $\alpha$, $\beta$, and $\gamma$ are the solutions of the following system of equations
\begin{equation}
	\left\{ \,
	\begin{IEEEeqnarraybox}[][c]{l?s}
	\IEEEstrut
	\alpha + \beta + \gamma = \left( 1 + \frac{k_f}{k_D} \right) \frac{\sqrt{D_A}}{a}, \\
	\alpha \gamma + \beta \gamma + \alpha \beta = k_b - k_d, \\
	\alpha \beta \gamma = k_b \frac{\sqrt{D_A}}{a} - k_d \left( 1 + \frac{k_f}{k_D} \right) \frac{\sqrt{D_A}}{a},
	\IEEEstrut
	\end{IEEEeqnarraybox}
	\right.
	\label{Eq. AlphaBetaGamma}
\end{equation} 
and $k_D = 4 \pi a D_A$.          
\end{theorem} 

\begin{IEEEproof}
Please refer to the Appendix.
\end{IEEEproof}

\textit{Remark 1:} $\alpha$, $\beta$, and $\gamma$ may be complex numbers. As a result, complex exponential and complementary error function have to be used for evaluation of $\w(\cdot, \cdot)$. However, the sum of the three $\w(\cdot, \cdot)$ terms on the right hand side of (\ref{Eq. Green's function}) is always a real number. 


\subsection{Channel Impulse Response}
Given the Green's function derived in the previous section, we can calculate the channel impulse response via (\ref{Eq. RelationshipReduced}). This leads to 
\begin{IEEEeqnarray}{rCl} 
	\label{Eq. Channel_Impulse_Response}
 \pac{t} & = & \frac{k_f e^{-k_dt}}{4\pi r_0 a \sqrt{D_A}} \left\lbrace \frac{\alpha \w\left( \frac{r_0 - a}{\sqrt{ 4 D_A t}}, \alpha \sqrt{t} \right)}{(\gamma - \alpha)(\alpha - \beta)}  \right. \nonumber \\
 && \hspace*{-2 mm} \left. +\> \frac{\beta  \w\left( \frac{r_0 - a}{\sqrt{4 D_A t}}, \beta \sqrt{t} \right)}{(\beta - \gamma)(\alpha - \beta)}  
  + \hspace*{-1 mm} \frac{\gamma \w\left( \frac{r_0 - a}{\sqrt{4 D_A t}}, \gamma \sqrt{t} \right)}{(\beta - \gamma)(\gamma - \alpha)} \right\rbrace. \nonumber \\*  	
\end{IEEEeqnarray} 

Finally, the number of $C$ molecules expected on the surface of the receiver after impulsive release of $N_A$ $A$ molecules at the transmitter can be obtained as 
\begin{IEEEeqnarray}{c} 
	\label{Eq. ExpectedReceSignal} 
	\overline{N}_C(t) = N_A \pac{t}.
\end{IEEEeqnarray}              
\section{simulation framework and results}
\label{Sec. Simulations} 
In this section, we first briefly describe our simulation framework. Then, we present simulation and analytical results for evaluation of the accuracy of the derived closed-form expression for the channel impulse response.

\subsection{Simulation Framework} 
We employ a Brownian motion particle-based simulation, where the precise locations of all individual molecules are
tracked throughout the simulation environment. In the adopted simulation algorithm, time is advanced in discrete steps of $\Delta t$ seconds. In order to jointly simulate the reactions, i.e., Eqs. (\ref{Eq.DegReaction}) and (\ref{Eq.RevReaction}), and the diffusion of the molecules, we combine in our simulator the algorithm proposed for the simulation of a first-order reaction in \cite{Steven_Andrews} with the algorithm for simulation of a second-order reversible reaction introduced in \cite{WoldeJ1}. In particular, in each step of the simulation, we perform the following operations:
 
$1$) Any $A$ molecule undergoes a random walk, where the new position of the molecule in each Cartesian coordinate is obtained by sampling a Gaussian random variable with mean $0$ and variance $\sqrt{2D_A\Delta t}$.

$2$) A uniformly distributed random number $l_1 \in [0,1]$ is generated for each $A$ molecule. Then, a given $A$ molecule degrades if its $l_1 \leq \pr \left(\text{Reaction} \, k_d \right)$, where $\pr \left(\text{Reaction} \, k_d \right)$ is the degradation probability of a given $A$ molecule in $\Delta t$ seconds, and is given by \cite[Eq. (13)]{Steven_Andrews} 
\begin{IEEEeqnarray}{c} 
	\label{Eq. ProbUniMolReac} 
	\pr \left(\text{Reaction} \, k_d \right) = 1 - \exp(-k_d \Delta t).
\end{IEEEeqnarray}

$3$) If the final position of an $A$ molecule at the end of a simulation step leads to an overlap with the receiver, a uniformly distributed random number $l_2 \in [0,1]$ is generated. Then, this displacement is accepted as an occurrence of the forward reaction if $l_2 \leq \pr \left( \text{Reaction} \, k_f \right)$. $\pr \left( \text{Reaction} \, k_f \right)$ is the probability of the forward reaction and given as \cite[Eq. (22)]{WoldeJ1} 
\begin{IEEEeqnarray}{c}
	\label{Eq. ProbBiMolReacForwardAcc} 
	\pr \left( \text{Reaction} \, k_f \right) = \frac{k_f \Delta t}{4\pi \rho},
\end{IEEEeqnarray}
where $\rho$ is a normalization factor that can be evaluated as 
\begin{IEEEeqnarray}{c} 
	\label{Eq. ForwardReacNormFac} 
	\rho = \int_{a}^{\infty} \Pr \left( \text{Ovr} | \vec{r}, \Delta t \right) r^2 \dif r.
\end{IEEEeqnarray}
Here, $\Pr \left( \text{Ovr} | \vec{r}, \Delta t \right)$ is the probability that a given $A$ molecule at position $\vec{r}$ \emph{overlaps} with the receiver in $\Delta t$ seconds, and can be written as \cite[Eq. (B3)]{WoldeJ1} 
\begin{IEEEeqnarray}{rCl}
	\label{Eq. ProbOverlap} 
	\Pr \left( \text{Ovr} | \vec{r}, \Delta t \right) & = & \frac{a}{2r \sqrt{\pi}} \left[ \exp\left(\frac{-(r+a)^2}{\sigma^2}\right) \right. \nonumber \\ 
	&& \hspace*{-27 mm} \left. -\>  \exp\left( \frac{-(r-a)^2}{\sigma^2} \right) \right] \hs + \hs \frac{1}{2} \left[ \erf\left( \frac{r+a}{\sigma} \right)  
	  + \erf\left( \frac{a-r}{\sigma} \right) \right], \nonumber \\*  
\end{IEEEeqnarray}
where $\sigma^2 = 4 D_A \Delta t$ and $\erf (\cdot)$ denotes the error function. By occurrence of the forward reaction, the overlapped $A$ molecule is removed and a new $C$ molecule is placed on the surface of the receiver at the position where the receiver surface intersects with a straight line describing the displacement of the $A$ molecule, i.e., the line between the positions of the molecule at the beginning and at the end of the simulation step. If $l_2 > \pr \left( \text{Reaction} \, k_f \right)$, then the overlapping $A$ molecule is returned to its previous position, i.e, the position it had at the beginning of the simulation step. 

$4$) For each $C$ molecule, a uniformly distributed random number $l_3 \in [0,1]$ is generated. Then, the backward reaction in (\ref{Eq.RevReaction}) occurs if $l_3 \leq \pr \left(\text{Reaction} \, k_b \right)$, where $\pr \left(\text{Reaction} \, k_b \right)$ is the probability that a given $C$ molecule on the surface of the receiver reverts back and produces an $A$ molecule outside the receiver. $\pr \left(\text{Reaction} \, k_b \right)$ can be evaluated via (\ref{Eq. ProbUniMolReac}) after substituting $k_d$ with $k_b$. The radial position of the new $A$ molecule is sampled from the normalized distribution $\Pr \left( \text{Ovr} | \vec{r}, \Delta t \right)r^2 / \rho $ and its angular coordinates, i.e., $\theta$ and $\phi$, are uniformly distributed, see \cite{WoldeJ1}.      
\subsection{Simulation Results}
In this subsection, we present simulation and analytical results for evaluation of the accuracy of the derived closed-form expression for the channel impulse response. 

In order to focus on the impact of the two \emph{chemical} reaction mechanisms introduced in Section \ref{Sec.SysMod}, i.e., the first order degradation reaction \eqref{Eq.DegReaction} and the second-order reversible reaction \eqref{Eq.RevReaction} on the surface of the receiver, on the characteristic of the received signal at the receiver, we keep the \emph{physical} parameters of the environment and the receiver constant throughout this subsection. In particular, we assume that $a = 0.5 \, \mu$m and $r_0 = 1 \, \mu$m. Furthermore, we assume that the diffusion coefficient of the information molecules is $D_A = 5 \times 10^{-9} \, \frac{\text{m}^2}{\text{s}}$ and $N_A = 5000$. The only parameters that we vary are $k_d$, $k_f$, and $k_b$. 

For all figures, the number of $C$ molecules expected on the surface of the receiver, i.e., $\overline{N}_C(t)$, was evaluated via \eqref{Eq. ExpectedReceSignal}. The simulation results were averaged over $5 \times 10^{4}$ independent releases of $N_A$ $A$ molecules at the transmitter and a simulation step size of $\Delta t = 0.5 \times  10^{-7}$s was chosen. We also note that the values of $k_f$, $k_b$, and $k_d$ were chosen such that the impact of changing any of these parameters can be observed over the time scale that is simulated.

\begin{figure}[!t] 
	\centering
	\hspace*{-0.5 cm}
	\includegraphics[scale = 0.34]{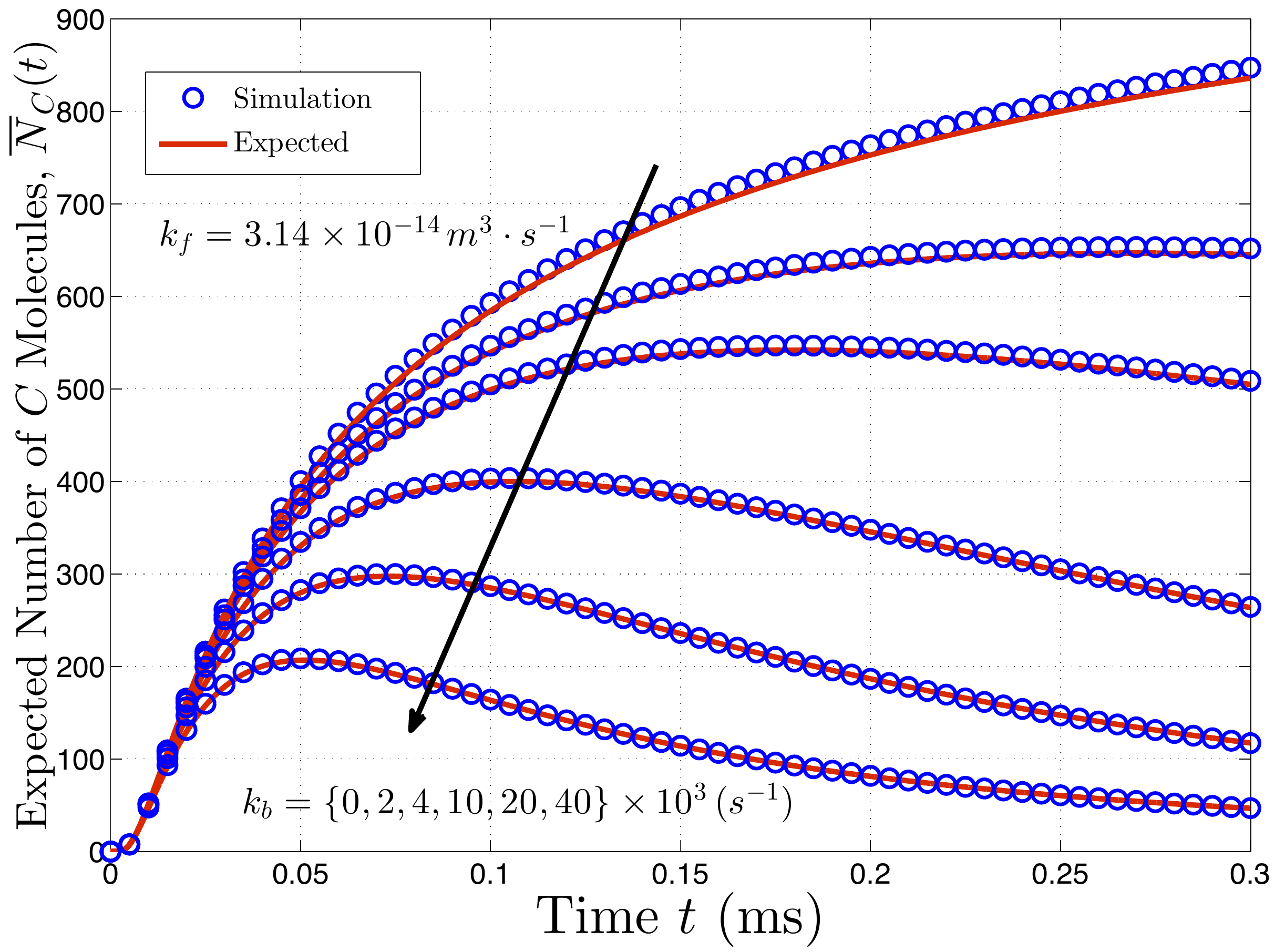}\vspace*{-4 mm}
	\caption{$\overline{N}_C(t)$ as a function of time $t$, when $k_d =0$.}\vspace*{-5 mm} 
	\label{Fig.ForwardBackward}
\end{figure} 
In Fig. \ref{Fig.ForwardBackward}, the number of $C$ molecules expected on the surface of the receiver, $\overline{N}_C(t)$, is shown as a function of time $t$ for system parameters $k_d = 0$ $\text{s}^{-1}$, $k_f = 3.14 \times 10^{-14}$ $\text{molecule}^{3}\cdot \text{s}^{-1}$, and $k_b = \{0, 2, 4, 10, 20, 40 \} \times 10^3$ $\text{s}^{-1}$. Fig. \ref{Fig.ForwardBackward} shows that when $k_b =0$, $\overline{N}_C(t)$ increases with increasing $t$ over the time scale that is simulated. This is because when $k_b = 0$, the $C$ molecules produced on the surface of the receiver cannot reverse back via the backward reaction in \eqref{Eq.RevReaction} and, as a result, stay on the surface of the receiver. However, when $k_b > 0$, for any $C$ molecule, there is a non-zero probability that it may reverse back and produce an $A$ molecule in the vicinity of the receiver surface. This new $A$ molecule may associate again with a $B$ molecules on the boundary of the receiver to produce a new $C$ molecule, or it may diffuse away from the receiver and not contribute to the production of a $C$ molecule. This is the reason why, when $k_b > 0$, $\overline{N}_C(t)$ eventually decreases with increasing $t$. We can also see that degradation occurs sooner and at a faster rate for larger $k_b$. This is because increasing $k_b$ increases the rate at which new $A$ molecules (produced by the backward reaction in \eqref{Eq.RevReaction}) escape from the vicinity of the receiver. Furthermore, we note the excellent match between simulation and analytical results.
 
\begin{figure}[!t]
	\centering
	\hspace*{-0.5 cm}
	\includegraphics[scale = 0.32]{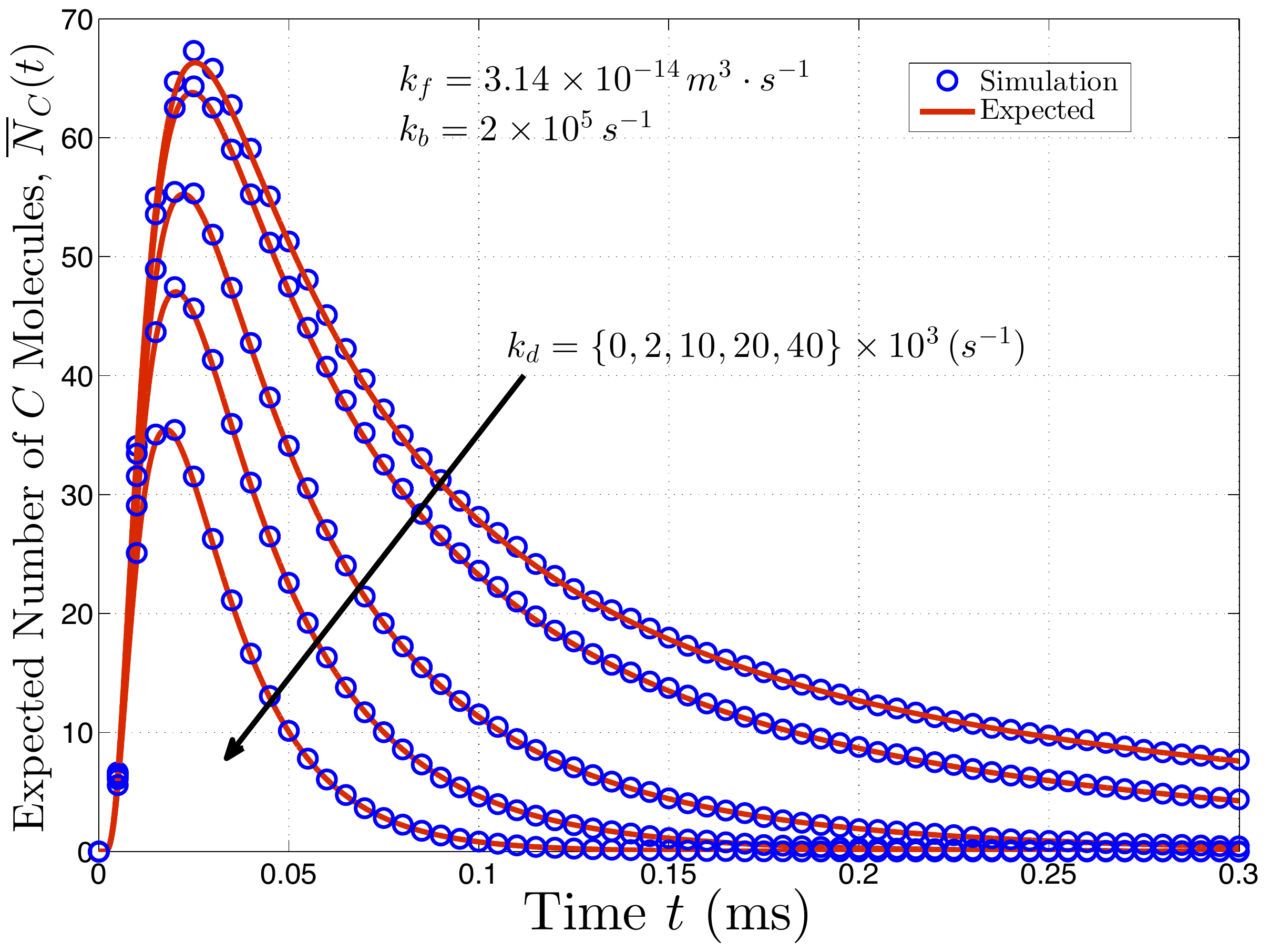}\vspace*{-4 mm}
	\caption{$\overline{N}_C(t)$ as a function of time $t$.}\vspace*{-5 mm} 
	\label{Fig.ForwardBackwardDeg}
\end{figure} 
In Fig. \ref{Fig.ForwardBackwardDeg}, $\overline{N}_C(t)$ is evaluated as a function of time $t$ for system parameters $k_f = 3.14 \times 10^{-14}$ $\text{molecule}^{3}\cdot \text{s}^{-1}$, $k_b = 2 \times 10^5$ $\text{s}^{-1}$, and $k_d = \{0, 2, 10, 20, 40 \} \times 10^3$ $\text{s}^{-1}$. In this figure, we keep the chemical parameters of the reversible reaction mechanism at the receiver constant to focus on the impact of the degradation reaction in the channel. As expected, $\overline{N}_C(t)$ decreases for increasing $k_d$. This is because larger values of $k_d$ increase the probability that a given $A$ molecule degrades in the channel and cannot produce a $C$ molecule at the receiver. As a result, fewer $A$ molecules contribute to the association reaction.                      
\section{conclusions} 
In this paper, we considered a diffusive molecular communication channel between a pair of transmitter and receiver nano-machines. We modelled the reception at the receiver as a second-order reaction mechanism, where information molecules released by the transmitter into a fluid environment could reversibly react with the receptor protein molecules covering the surface of the receiver. We furthermore assumed that the information molecules can degrade in the channel via a first-order degradation reaction. We derived a closed-form time domain expression for the channel impulse response of the system and verified its accuracy via particle-based simulations. An excellent match between analytical and simulation results was observed.    
\label{Sec.Con}  
\appendix[Proof of Theorem 1]
We denote the Fourier, inverse Fourier, Laplace, and inverse Laplace transforms by $F\{\cdot\}$, $F^{-1}\{\cdot\}$, $L\{\cdot\}$, and $L^{-1}\{\cdot\}$, respectively.

We start by solving reaction-diffusion equation (\ref{Eq. Reaction-DiffusionU}) for initial condition (\ref{Eq. InitialConditionU}). Taking the Fourier transform of (\ref{Eq. Reaction-DiffusionU}) with respect to $r$ leads to the following partial differential equation 
\begin{IEEEeqnarray}{c}
	\label{Eq. AppFourierPartial} 
	\frac{\partial \overline{U}(\kappa,t | r_0)}{\partial t} = -(D_A^2 \kappa^2 + k_d ) \overline{U}(\kappa,t | r_0) 
 \end{IEEEeqnarray}
where $\overline{U}(\kappa,t | r_0) = F\{rU(r,t|r_0)\}$, with $\kappa$ representing the Fourier domain variable, is given by 
\begin{IEEEeqnarray}{c} 
	\label{Eq. AppFourierTransform}
	\overline{U}(\kappa,t | r_0) = \frac{1}{2\pi} \int_{- \infty}^{+ \infty} rU(r,t|r_0) \exp(-j\kappa r) \dif r, 
\end{IEEEeqnarray}
 
Solving \eqref{Eq. AppFourierPartial} for $\overline{U}(\kappa,t | r_0)$ leads to 
\begin{IEEEeqnarray}{c} 
	\label{Eq. App1} 
	\overline{U}(\kappa,t | r_0) = C_0 \exp(-D_A \kappa^2 t + k_dt),
\end{IEEEeqnarray} 
where $C_0$ is a constant to be determined by the initial condition. Taking the Fourier transform of \eqref{Eq. InitialConditionU} and employing (\ref{Eq. App1}), constant $C_0$ is obtained as 
\begin{IEEEeqnarray}{c}
	\label{Eq. AppConstantC} 
	C_0 = \frac{\exp(-j\kappa r_0)}{4 \pi r_0}.
\end{IEEEeqnarray}

Finally, $rU(r,t|r_0)$ can be evaluated as 
\begin{IEEEeqnarray}{rCl}
	\label{Eq. AppU_time} 
	rU(r,t|r_0) & = & F^{-1}\{\overline{U}(\kappa,t | r_0)\} \nonumber \\
	& = & \frac{\exp(-k_d t)}{8 \pi r_0 \sqrt{\pi D_A t}} \exp \left( \frac{-(r - r_0)^2}{4 D_A t} \right). 
\end{IEEEeqnarray} 

Next, we solve reaction-diffusion equation (\ref{Eq. Reaction-DiffusionV}) for initial condition (\ref{Eq. ConditionV}). In order to do so, we first apply the Laplace transform with respect to $t$ to (\ref{Eq. Reaction-DiffusionV}) which results in 
\begin{IEEEeqnarray}{rCl}
	\label{Eq. App2} 
	sr\overline{V}(r,s|r_0) - rV(r,t = 0 |r_0) & = & D_A \frac{\partial^2}{\partial r^2} \big(r \overline{V}(r,s|r_0) \big) \nonumber \\ 
	&& -\> k_d r \overline{V}(r,s|r_0),
\end{IEEEeqnarray}
where $\overline{V}(r,s|r_0) = L\{V(r,t|r_0)\}$, i.e., 
\begin{IEEEeqnarray}{c}
	\label{Eq. AppLaplaceTransform} 
	\overline{V}(r,s|r_0) = \int_{0}^{\infty} V(r,t|r_0) \exp(-st) \dif t. 
\end{IEEEeqnarray}

The second term on the left hand side of (\ref{Eq. App2}) is zero (see (\ref{Eq. ConditionV})). Eq. (\ref{Eq. App2}) is a partial differential equation of function $r\overline{V}(r,s|r_0)$. As can be seen from \eqref{Eq. AppU_time}, $U(r \to \infty,t | r_0) =0$. Thus, in order to satisfy (\ref{Eq. BounCon1}), $V(r \to \infty, t | r_0) = 0$. Taking this into account, the solution of \eqref{Eq. App2} can be written as 
\begin{IEEEeqnarray}{c}
	\label{Eq. AppV_L} 
	\overline{V}(r,s|r_0) = \frac{q}{r} \exp\left( - r \sqrt{\frac{s + k_d}{D_A}} \right),
\end{IEEEeqnarray}
where $q$ is a constant that can be used to ensure that $U(r,t | r_0)$ and $V(r,t | r_0)$ jointly satisfy boundary condition (\ref{Eq. BounCon2}). In order to find $q$, we first evaluate $L\{P_A(r,t | r_0)\} = \overline{P}_A(r,s | r_0) = L\{U(r,t | r_0)\} + L\{V(r,t| r_0)\}$ which yields 
\begin{IEEEeqnarray}{rCl} 
 \label{Eq. AppCompleteSolution_L} 
 \overline{P}_A(r,s | r_0) & = & \frac{\exp\left( - \sqrt{\frac{s + k_d}{D_A}} ( r - r_0 ) \right)}{8 \pi r r_0 \sqrt{D_A (s + k_d)}} \nonumber \\
 && +\> \frac{q}{r} \exp\left( - r \sqrt{\frac{s + k_d}{D_A}} \right),
\end{IEEEeqnarray} 
where we used \cite[Eq. 29.3.84]{abramowitz} 
\begin{IEEEeqnarray}{c} 
	L \left\lbrace \frac{1}{\sqrt{\pi t}} \exp\left( \frac{-b^2}{4t} \right) \right\rbrace = \frac{\exp(-b \sqrt{s})}{\sqrt{s}}, 
\end{IEEEeqnarray}
for evaluation of $L\{U(r,t | r_0)\}$. Then, we calculate the Laplace transform of boundary condition (\ref{Eq. BounCon2}), which leads to 
\begin{IEEEeqnarray}{c} 
	\label{Eq. AppBoundaryCon2_L} 
	\frac{\partial \overline{P}_A(r,s | r_0)}{\partial r} \bigg|_{r = a} = \frac{k_f s}{a k_D s + k_b a k_D} \overline{P}_A(r,s | r_0). 
\end{IEEEeqnarray}

Now, taking the derivative of (\ref{Eq. AppCompleteSolution_L}) with respect to $r$, substituting the resulting equation into (\ref{Eq. AppBoundaryCon2_L}), and solving the corresponding equation for $q$ yields 
\begin{IEEEeqnarray}{rCl} 
	\label{Eq. AppConstantq} 
	q & = & \frac{a k_D \sqrt{\frac{s+k_d}{D_A}} - k_D - \frac{k_f s}{s + k_b}}{a k_D \sqrt{\frac{s+k_d}{D_A}} + k_D + \frac{k_f s}{s + k_b}} \times \frac{1}{8 \pi r_0 \sqrt{D_A(s + k_d)}} \nonumber \\
	&& \times \> \exp\left( - \sqrt{\frac{s+k_d}{D_A}} (r_0 - 2a) \right).    
\end{IEEEeqnarray} 

The Laplace transform of the final solution can be evaluated by substituting (\ref{Eq. AppConstantq}) into (\ref{Eq. AppCompleteSolution_L}), and can be written as the sum of three terms as follows:
\begin{IEEEeqnarray}{lCl} 
	\label{Eq. AppCompleteSolution_L_and_q} 
	 \overline{P}_A(r,s | r_0)  & = & \frac{\exp\left( - \sqrt{\frac{s + k_d}{D_A}} ( r - r_0 ) \right)}{8 \pi r r_0 \sqrt{D_A (s + k_d)}} \nonumber \\ 
	&& +\> \frac{\exp\left( - \sqrt{\frac{s + k_d}{D_A}} ( r + r_0 -2a ) \right)}{8 \pi r r_0 \sqrt{D_A (s + k_d)}} \nonumber \\  
	&& -\> \left[ \frac{k_D + \frac{s k_f}{s + k_b}}{a k_D \sqrt{\frac{s + k_d}{D_A}} + k_D + \frac{s k_f}{ s + k_b}} \right. \nonumber \\ 
	&& \left. \times\> \frac{\exp\left( - \sqrt{\frac{s + k_d}{D_A}} ( r + r_0 -2a ) \right)}{4 \pi r r_0 \sqrt{D_A(s + k_d)}} \right].
\end{IEEEeqnarray}

Taking the inverse Laplace transform of the first two terms on the right hand side of (\ref{Eq. AppCompleteSolution_L_and_q}) yields the first two terms on the right hand side of (\ref{Eq. Green's function}). The denominator of the third term on the right hand side in (\ref{Eq. AppCompleteSolution_L_and_q}) can be rearranged and written as 
\begin{IEEEeqnarray}{c} 
	\label{Eq. AppDenominator} 
	\left[ (s \hs + \hs k_d)^{3/2} + (k_b - k_d)\sqrt{s \hs + \hs k_d} - \frac{k_D \hs + \hs k_f}{a k_D} \sqrt{D_A} k_d \right. \nonumber \\ 
	\left. + \frac{k_D \hs + \hs k_f}{a k_D} \sqrt{D_A} (s \hs + \hs k_d) + \frac{k_b}{a} \sqrt{D_A} \right] \left(4 \pi r r_0 \sqrt{D_A(s \hs + \hs k_d)} \right). \nonumber \\*
\end{IEEEeqnarray} 

The terms in brackets can be interpreted as a cubic equation in $\sqrt{s+k_d}$. Let us assume that $-\alpha$, $-\beta$, and $-\gamma$ are the roots of this cubic equation. Then, it can be easily verified that these roots have to satisfy the system of equations in (\ref{Eq. AlphaBetaGamma}). Employing the partial fraction expansion technique, the third term on the right hand side of (\ref{Eq. AppCompleteSolution_L_and_q}) can be written as 
\begin{IEEEeqnarray}{c} 
	\label{Eq. AppLastTermDecompose} 
	\left( \frac{\eta_1}{\alpha + \sqrt{s + k_d}} + \frac{\eta_2}{\beta + \sqrt{s + k_d}} + \frac{\eta_3}{\gamma + \sqrt{s + k_d}} \right)  \nonumber\\ 
	  \times \frac{1}{4 \pi r r_0 \sqrt{D_A(s + k_d)}} \exp\left( - \sqrt{\frac{s + k_d}{D_A}} ( r + r_0 -2a ) \right), \nonumber \\*
\end{IEEEeqnarray}
where $\eta_1$, $\eta_2$, and $\eta_3$ are the residues and given by (\ref{Eq. eta1})-(\ref{Eq. eta3}). Taking the inverse Laplace transform of (\ref{Eq. AppLastTermDecompose}) results in the last three terms on the right hand side of (\ref{Eq. Green's function}), where we used \cite[Eq. 29.3.90]{abramowitz} 
\begin{IEEEeqnarray}{rCl} 
	\label{Eq. AppLaplaceInverseLastStep} 
	L^{-1}\left\lbrace \frac{\exp(-n\sqrt{s})}{\sqrt{s}(m + \sqrt{s})} \right\rbrace & = & \exp\left( nm + m^2 t \right) \nonumber \\ 
	 && +\> \erfc\left( \frac{n}{2\sqrt{t}} + m\sqrt{t} \right).
\end{IEEEeqnarray}           
\bibliography{IEEEabrv,Library}
\end{document}